\documentclass{vhepu14}

\usepackage{amsbsy}
\usepackage{multirow}
\usepackage{amsmath}

\def\be{\begin{equation}}
\def\ee{\end{equation}}
\def\bes{\begin{equation*}}
\def\ees{\end{equation*}}
\def\bea{\begin{eqnarray}}
\def\eea{\end{eqnarray}}

\newcommand{\Photo}{}

\begin{document}
\vspace*{4cm}
\title{THE RADIO SIGNAL FROM EXTENSIVE AIR SHOWERS}

\author{BENO\^IT REVENU}

\address{Subatech, \'Ecole des Mines de Nantes, CNRS/IN2P3, Universit\'e de Nantes\\
	4 rue Alfred Kastler, BP20722, 44307 Nantes C\'EDEX 3, FRANCE}

\maketitle\abstracts{
The field of ultra-high energy cosmic rays made a lot of progresses last years with large area experiments such as the Pierre Auger Observatory, HiRes and the Telescope Array. A suppression of the cosmic ray flux at energies above $5.5\times 10^{19}$~eV is observed at a very high level of significance but the origin of this cut-off is not established: it can be due to the Greisen-Zatsepin-Kuzmin suppression but it can also reflect the upper limit of particle acceleration in astrophysical objects. The key characteristics to be measured on cosmic rays is their composition. Upper limits are set above $10^{18}$~eV on primary photons and neutrinos and primary cosmic rays are expected to be hadrons. Identifying the precise composition (light or heavy nuclei) will permit to solve the puzzle. It has been proven that the radio signal emitted by the extensive air showers initiated by ultra-high energy cosmic rays reflects their longitudinal profile and can help in constraining the primary particle. We review in this paper the emission mechanisms as a function of the frequency of the electric field.
}

\section{Observing extensive air showers}
In the current large area ultra-high energy cosmic rays experiments (Auger~\cite{Allekotte:2007sf,fdpao2010} and TA~\cite{AbuZayyad201287,Tinyakov201429}), extensive air showers are studied through the secondary particles reaching the ground level and by collecting the fluorescence light emitted by the atmospheric diazote molecules excited by the passage of the charged particles. Surface detectors (SD) arrays are composed of particle detectors (Cerenkov water tanks, scintillators...) having a high duty cycle (around $100$~\%) and that allow to estimate the arrival direction of the shower and its core position. During dark nights (duty cycle of $14$~\% in Auger), the fluorescence detectors (FD) detect the longitudinal profile of the showers through a calorimetric measurement during their development in the air. A FD directly measures the shower energy and estimates the nature of the primary cosmic ray by the determination of the atmospheric depth $X_{\mathrm{max}}$ corresponding to the maximum number of secondary particles in the shower. Hybrid events, detected by both SD and FD, allow to calibrate the SD using the FD reconstruction of the energy. The need for a better event-by-event identification of the nature of the primary particle have speed up the research and development of additionnal detection techniques such as the measurement of the radio signal emitted by the electrons and positrons of the showers.

\section{Electric field from air showers}
The radio signal associated to cosmic rays was detected for the first time in 1965\cite{jelley1965} but the detailed characteristics of the electric field have been understood in the last years using the data from several experiments (CODALEMA~\cite{codalema2005}, LOPES~\cite{falcke2005}, LOFAR~\cite{nelleslofaricrc2013}, AERA~\cite{neuser-arena-2014}...). It has been shown that the emission is coherent for wavelengths larger than or of the order of the typical dimensions of air showers: the scale of their longitudinal development ($10$~km, $30$~kHz), their lateral spread ($1$~km, $0.3$~MHz) and the thickness of their front ($10$~m, $30$~MHz). The individual electric fields emitted by all high energy ($>80$~MeV) electrons and positrons of air shower (we neglect the emission from muons and ions, being much heavier) interfere with each other: if the observation wavelength is larger than the size of the emitting region, the fields add-up coherently so that the total electric field is proportionnal to the number of electrons and positrons, and then, to the primary energy. On the contrary, if the wavelength is smaller than the size of the emitting region, the fields interfere destructively. We should therefore expect a cut-off frequency in the full frequency spectrum. The radio signal has a clear coherent contribution over a large frequency range: from tens of kHz up to some hundreds of MHz. It has been shown that the signal in the frequency range 30-80~MHz allows to reconstruct precisely the incoming direction of the shower (angular resolution below $1^\circ$, see~\cite{revenuICRC2011}).  The deep understanding of the radio signal we reached today allows to use simulations to estimate the $X_{\mathrm{max}}$ and LOFAR~\cite{buitinklofaricrc2013} obtained a resolution on this parameter of $20~\mathrm{g}~\mathrm{cm}^{-2}$, similar to the resolution achieved by the FD. The energy estimation is underway, for instance in the AERA experiment.

\subsection{Emission mechanisms}
In 1967, Kahn and Lerche~\cite{kl1966} studied the dominant mechanism: the emission from the time varying transverse currents induced by the interaction between the geomagnetic field and the charged particles (hereafter the geomagnetic mechanism). The systematic opposite drift of electrons and positrons when they propagate in the geomagnetic field associated to the interactions with the medium results into an average transverse current (with respect to the shower axis). This current induces a macroscopic electric field linearly polarized along the $\mathbf{a}\times\mathbf{B}$ direction, where $\mathbf{a}$ is the direction of the shower axis and $\mathbf{B}$ the direction of the geomagnetic field. The orientation of this electric field is independent on the observer's location.

A secondary emission process is due to the excess of electrons with respect to positrons in air showers (Askaryan~\cite{ask1962}). This net negative charge excess is due to the fast in-flight positrons annihilation and to direct enrichment in electrons extracted from the medium by the Compton, Bhabha and Moeller diffusions. The charge-excess mechanism leads to a radial polarization pattern of the electric field in the plane transverse to the shower axis.

\begin{figure}[!ht]
\begin{center}
\begin{minipage}{0.3\linewidth}
\centerline{\includegraphics[scale=0.15,draft=false]{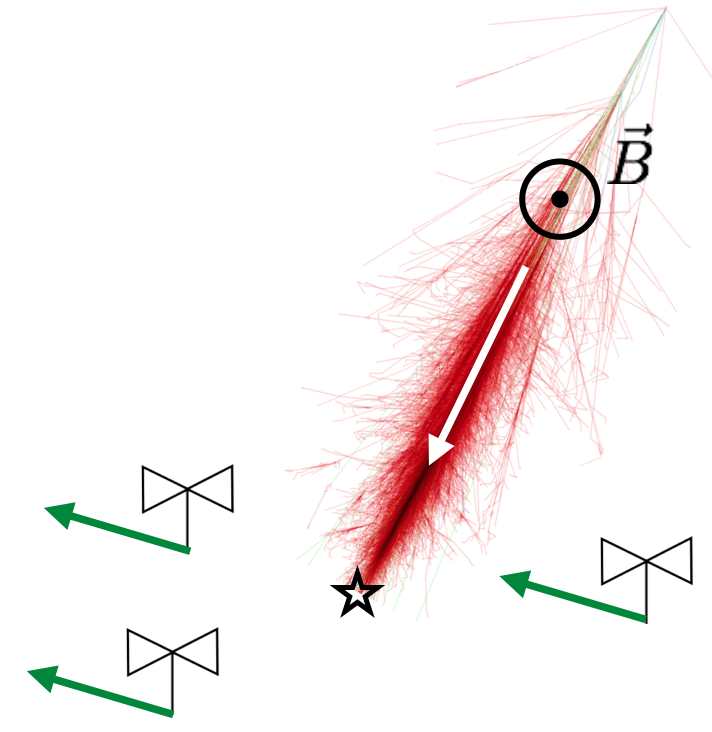}}
\end{minipage}\hfill
\begin{minipage}{0.3\linewidth}
\centerline{\includegraphics[scale=0.15,draft=false]{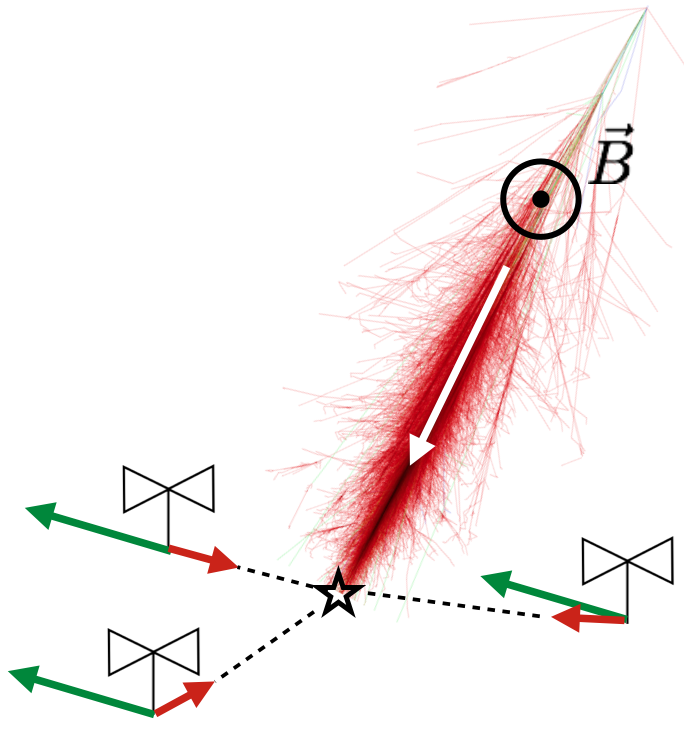}}
\end{minipage}\hfill
\begin{minipage}{0.3\linewidth}
\centerline{\includegraphics[scale=0.15,draft=false]{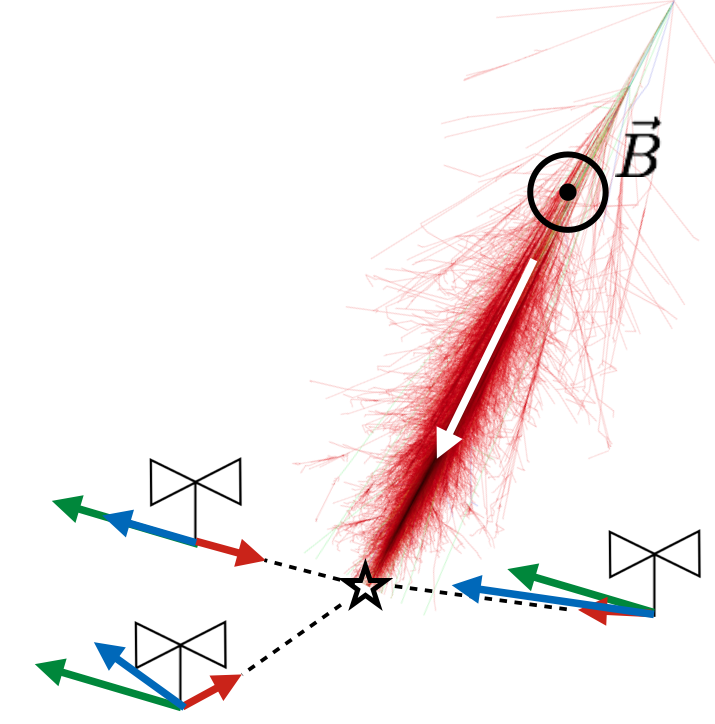}}
\end{minipage}
\caption[]{Left: in green, electric field generated by the geomagnetic mechanism. The polarization is linear and the same for all observers. Center: in red, electric field generated by the charge-excess mechanism. The polarization is linear with a radial pattern with respect to the shower core in the plane transverse to the shower axis. Right: in blue, total electric field measured by an observer. Its value and direction depends on the observer's location.}\label{schemegeo}
\end{center}
\end{figure}

An observer will detect the superposition of both electric fields. These fields depend on time through the varying number of electrons and positrons in the showers during its development in the air.
This superposition leads to an asymetry in the total electric field strength according to the observer's location with respect to the shower core and explains the departure from the radially symetric lateral distribution function proposed by Allan~\cite{allan1971} in 1971. The electric field value cannot be simply modelized by a 1D lateral function.
Figure~\ref{schemegeo} presents the superposition of both mechanisms.
Note that other mechanisms can also lead to the emission of electric fields: the presence of electric dipoles in the shower, the emission from the ions that are left behind the shower, the individual geosynchrotron emission but all of them are largely subdominant.

\subsection{Simulation approaches}\label{simu}
The simulation of the electric field emitted by air showers recently reached a mature and robust state. Different approaches were chosen in the past, using microscopic or macroscopic descriptions, in the time or frequency domains. The codes MGMR~\cite{mgmr2010} and EVA~\cite{eva2012} use the macroscopic transverse current as the main ingredient to the Maxwell's equations. The code SELFAS~\cite{selfas2012,marinARENA2012} uses the energy, impulsion, angular, position distributions of electrons and positrons from the shower universality description and computes the resulting electric field as the summation of all individual contributions (see Eq.~\ref{seleq}). The code ZHAireS~\cite{zhaires2012} uses the ZHS formalism and computes the electric field in both time and frequency domains. CoREAS~\cite{coreas2013} uses the end-point formalism and computes all individual electric fields directly inside a full CORSIKA simulated shower.
As an example, the following formula used in SELFAS gives the electric field emitted by a single particle of charge $q$ and lifetime $\tau=t_2-t_1$ (its charge $q$ is taken as $0$ before $t_1$ and after $t_2$):
\begin{equation}\label{seleq}
\mathbf{E}(\mathbf{r},t)=\frac{1}{4\pi\varepsilon_0}\left(\frac{q\,\mathbf{n}}{R^2(1-\eta\,\boldsymbol{\beta}\cdot\mathbf{n})}+\frac{1}{c}\frac{\partial}{\partial t}\frac{q\,\mathbf{n}}{R(1-\eta\, \boldsymbol{\beta}\cdot\mathbf{n})} -\frac{1}{c}\frac{\partial}{\partial t}\frac{q\,\boldsymbol{\beta}}{R(1-\eta\, \boldsymbol{\beta}\cdot\mathbf{n})}\right)_\mathrm{ret}
\end{equation}
where $\mathbf{r}$ is the observer's position, $\mathbf{n}$ is the unit vector between the particle and the observer, $\eta$ is the air refractive index, $\boldsymbol{\beta}$ the particle velocity and $R$ the distance between the particle and the observer. The field is evalutated at time $t_\mathrm{ret}=t-\eta\,R/c$. We obtain the complete electric field after summation over all electrons and positrons of the shower. The first term corresponds to the Coulombian contribution; it is negligible with respect to the two others. The global contribution of the second term for a system with equal numbers of electrons and positrons vanishes. Due to the net excess of electrons in air showers, this term is not negligible and constitutes the charge excess contribution. The third term contains the particle velocity and its time derivative is proportionnal to the Lorentz force: this is the geomagnetic term.
The particle velocity direction is close, at first order, to the shower axis direction. We note the presence of the air refractive index~$\eta$ leading to a boost of the electric field when the angle between the particle velocity and the line of sight is close to the Cherenkov angle. This is not actual Cherenkov radiation but this effect is important for observers located close to the shower axis. For these observers, in the time domain, the electric field pulse is sharpened making it dominant at higher frequencies (above $100$~MHz) and still important below $100$~MHz and this explains the flat electric profiles observed by LOPES~\cite{lopes2009ci} and CODALEMA in the past. The electric field is amplified along a ring of radius $R_C=\ell\sin\theta_{C}\sim 140$~m, $\ell\sim 6$~km being is the typical distance from the shower's maximum for $\sim 10^{18}$~eV primary energies and $\theta_C=\arccos(1/\eta)\sim 1.4^\circ$ is the Cherenkov angle in air. Observation of this ring would be of great interest because it probes the distance to the source of the electric field and therefore could help in differentiating proton or iron initiated showers. The ring radius being small, it would be necessary to detect shower with a dense array of radio stations which is not a good solution (not cheap).

\section{Radio signal in the range $\mathbf{30-200}$~MHz}
The frequency band $30-200$~MHz is intensively studied since the beginning of years 2000. It has been established that the main emission mechanism in this band is the geomagnetic contribution. Beside this contribution, the CODALEMA and AERA experiments observed the secondary electric field emitted by the charge excess mechanism using different observables.
The CODALEMA experiment compared the shower core positions obtained using the data from the radio array to the core positions obtained using the particle array. A significant statistical discrepancy between the two core positions was found: on average, the radio core positions is shifted by $\sim 25$~m to the east with respect to the particle core position. The radio core position is estimated using a radio lateral distribution function (LDF) of the electric field amplitude depending only on the distance to the shower axis; {\em ie} a radially symmetric LDF. Using SELFAS simulations including only the geomagnetic contribution, it has been shown that the core positions estimated from the radio and particle arrays were in good agreement. Using SELFAS with both mechanisms permitted to reproduced the same shift toward the east. This was the first strong indication of the presence of the charge excess mechanism in the data~\cite{marinICRC2011}.

\begin{figure}[!ht]
\begin{center}
\begin{minipage}{0.4\linewidth}
\centerline{\includegraphics[scale=0.3,draft=false]{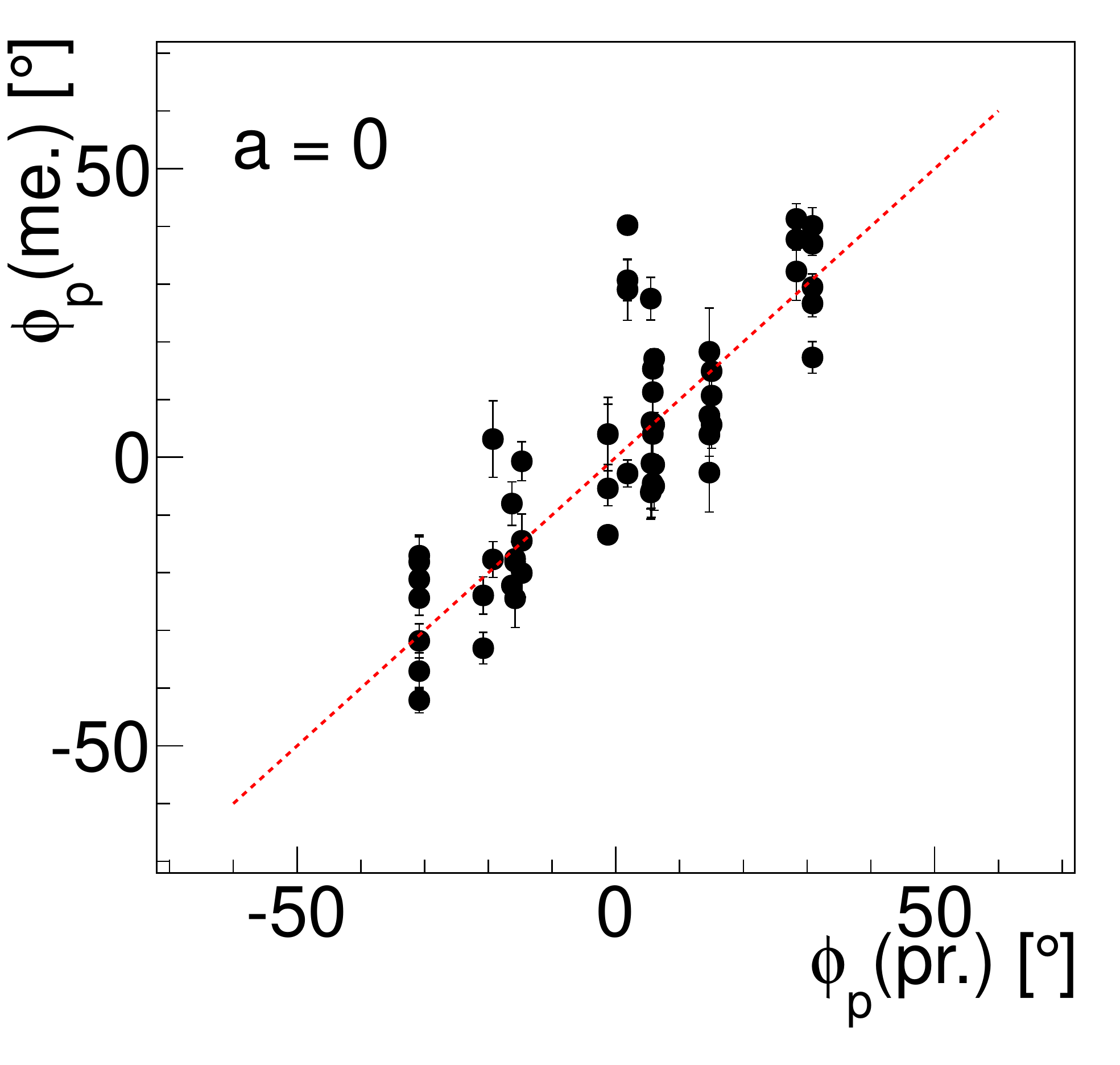}}
\end{minipage}
\begin{minipage}{0.4\linewidth}
\centerline{\includegraphics[scale=0.3,draft=false]{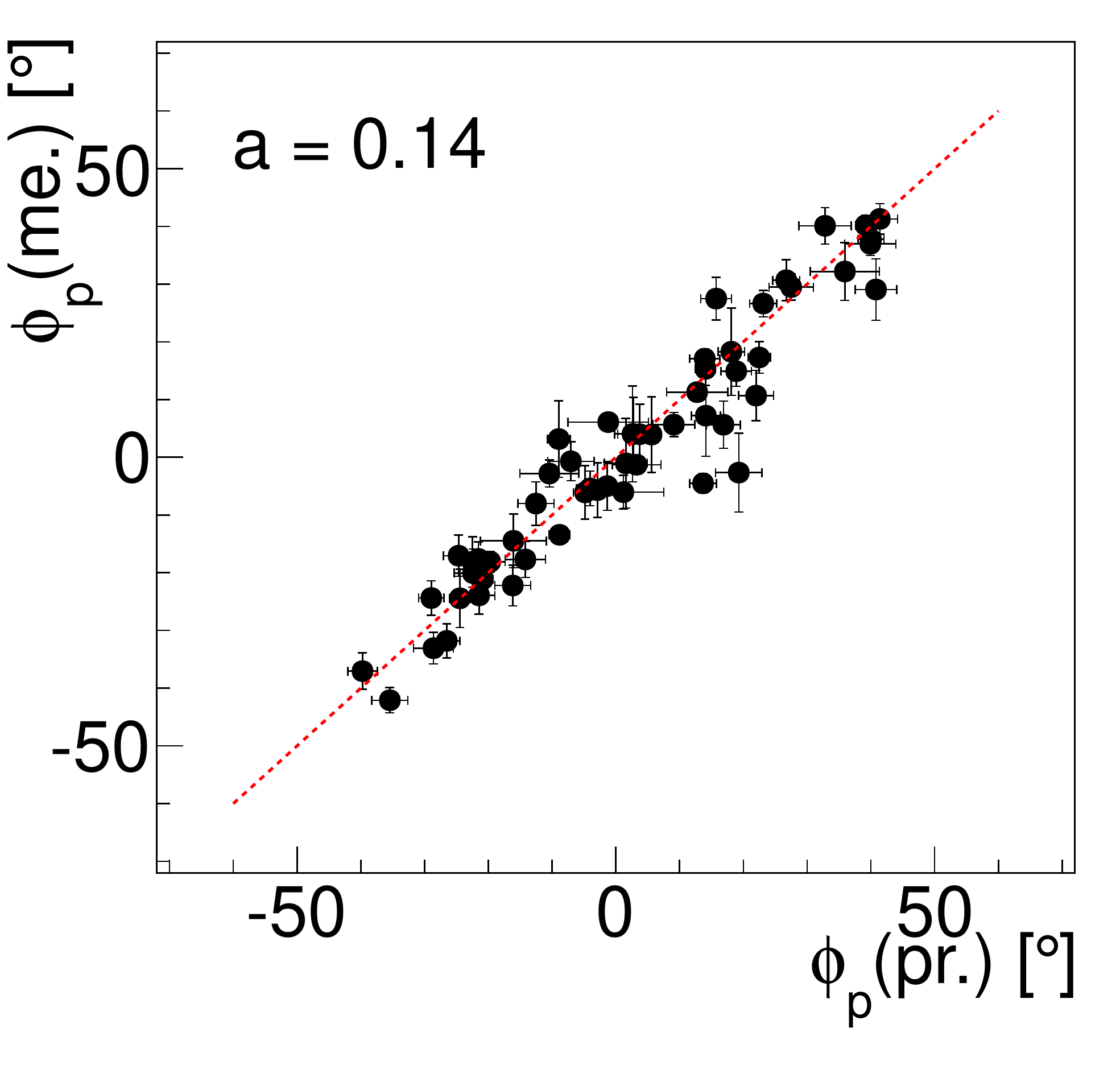}}
\end{minipage}
\caption[]{Left: correlation between the measured polarization angle and the predicted polarization angle in the case of a pure geomagnetic contribution. Right: same than the left figure but adding the contribution of the charge excess mechanism.}\label{polaera}
\end{center}
\end{figure}

The AERA experiment used the polarization information to exhibit the charge excess contribution. Following Figure~\ref{schemegeo}, an observer measures the superposition of two electric fields with different polarization patterns. If we consider only the geomagnetic mechanism, we can compute the polarization angle $\phi_P(\mathrm{pr.})$ (with respect to the geographic east) in the horizontal plane using the north-south (NS) and east-west (EW) components of the predicted electric field:
\bes
\phi_P(\mathrm{pr.})=\arctan\frac{(\boldsymbol{a}\times\mathbf{B})_\mathrm{NS}}{(\boldsymbol{a}\times\mathbf{B})_\mathrm{EW}}
\ees
Using the data, we can compute for each radio station $i$ in a given event the measured polarization angle in the horizontal plane (the antennas used in AERA do measure the electric field in both EW and NS directions). This angle $\phi^i_P(\mathrm{me.})$ is given by:
\bes
\phi^i_P(\mathrm{me.})=\arctan\frac{\varepsilon_\mathrm{NS}}{\varepsilon_\mathrm{EW}}
\ees
where $\varepsilon$ is the electric field amplitude in both EW and NS directions. The correlation between $\phi_P(\mathrm{pr.})$ and $\phi^i_P(\mathrm{me.})$ in the case of a pure geomagnetic contribution is presented in Figure~\ref{polaera}(left). For a given event with a known incoming direction ({\em ie} we know the axis direction $\mathbf{a}$) involving $N$ radio stations, we have a single value for $\phi_P(\mathrm{pr.})$ but we have $N$ values of the measured polarization angles $\phi^i_P(\mathrm{me.})$. This explains the bunches of different values of $\phi^i_P(\mathrm{me.})$ at a given value of $\phi_P(\mathrm{pr.})$.
The correlation is very clear: the Pearson correlation coefficient is $\rho_P=0.82^{+0.06}_{-0.04}$ at 95\% CL (with $\chi^2/\mathrm{ndf}=27$) which demonstrates that the geomagnetic is indeed dominant.
As we know that the charge excess mechanism is also there, we can refine our predicted value of the polarization angles $\phi_P(\mathrm{pr.})$. As discussed before, the electric field from this mechanism has a radial linar polarization in the plane transverse to the shower axis. The predicted polarization angle at each radio station $i$ can be deduced using the following formula:
\bes
\phi^i_P(\mathrm{me.})=\arctan
    \left(\frac
    {\sin\phi_{\mathrm{G}}\, \sin\alpha + a\, \sin\phi_{\mathrm{A}}^i}
    {\cos\phi_{\mathrm{G}}\, \sin\alpha + a\, \cos\phi_{\mathrm{A}}^i  }\right),
   \ees
   where $\phi_{\mathrm{G}}$ is the azimuth of the geomagnetic contribution, $\alpha$ the angle between the shower axis $\mathbf{a}$ and the geomagnetic field $\mathbf{B}$ and $\phi_{\mathrm{A}}^i$ is the azimuth of the charge excess contribution at the location of the radio station~$i$. The parameter $a$ is the ratio of the amplitude of the electric field from the charge excess contribution $\varepsilon_A$ to the amplitude of the electric field from the geomagnetic contribution $\varepsilon_G$, modulated by $\sin\alpha$ (in order to cancel the geomagnetic contribution when the shower incoming direction is parallel to the geomagnetic field):
   \bes
   a=\frac{|\varepsilon_A|}{|\varepsilon_G|}\sin\alpha
   \ees
  The average value of $a$ was computed using showers detected by AERA and we obtained: $a=0.14\pm 0.02$ (see~\cite{polarisationAERA2014}). This value should not be considered as the fraction of charge excess in the shower, as it depends also on the angular distance between the shower axis and the geomagnetic field through~$\sin\alpha$. Using this value for $a$, we can finally compute the predicted polarization angle $\phi^i_P(\mathrm{me.})$ for each radio station $i$ involved in a given event, taking into account both geomagnetic and charge excess contributions. The correlation between $\phi^i_P(\mathrm{me.})$
  and $\phi_P(\mathrm{me.})$ is presented in Figure~\ref{polaera}(right). The Pearson correlation coefficient rises up to $\rho_P=0.93^{+0.04}_{-0.03}$ at 95\% CL (with $\chi^2/\mathrm{ndf}=2.2$). The agreement between the data and the model is much better when considering both mechanisms.

In conclusion, the radio signal in the range $30-200$~MHz is very well described by the superposition of two mechanisms of electric field emission: the geomagnetic and the charge excess contributions.

\section{Radio signal in the GHz domain}

\begin{figure}[!ht]
\begin{center}
\begin{minipage}{0.4\linewidth}
\includegraphics[width=7cm,draft=false]{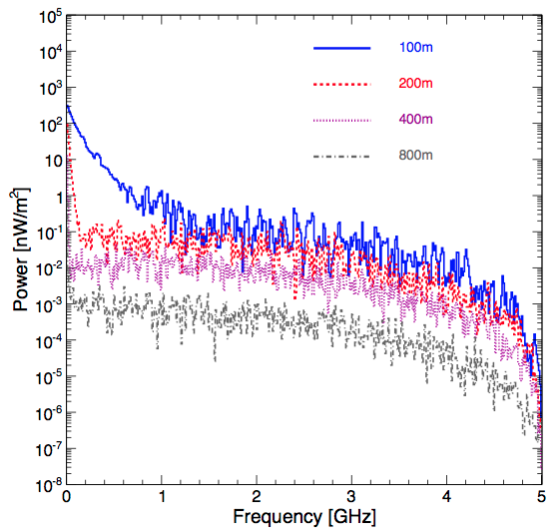}
\end{minipage}
\hspace{1cm}
\begin{minipage}{0.4\linewidth}
\includegraphics[width=7cm,draft=false]{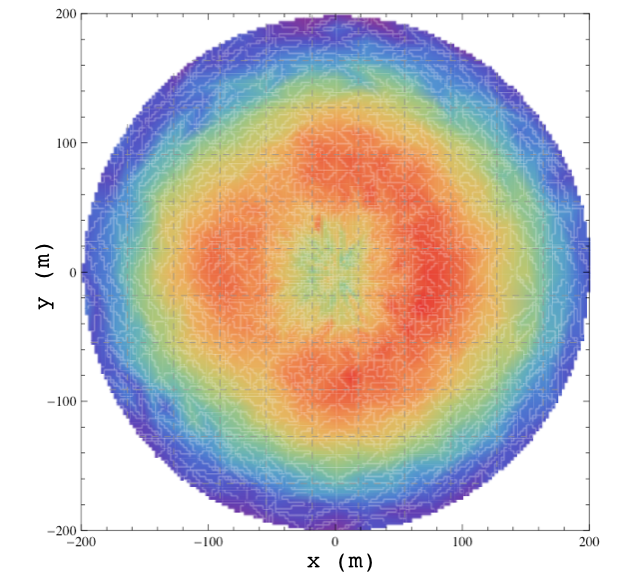}
\end{minipage}
\caption{Contribution at high frequencies of the electric field from the geomagnetic and charge excess mechanisms. Left: electric field power as a function of frequency. The coherent part lies at smaller frequencies as can be seen for antennas close to the shower axis ($100$~m and $200$~m). Right: amplitude of the same electric field between $300$~MHz and $1.2$~GHz a a function of the observer location at the groud level. Both plots were obtained for a simulated vertical shower (using SELFAS) initiated by an iron nucleus at $10^{17}$~eV.}\label{selfas_allf}
\end{center}
\end{figure}

The first signal we can think about in this frequency region is the GHz counterpart of the usual signal made of the geomagnetic and charge excess mechanisms, enhanced by the Cherenkov-like effect due to the air index effect as discussed in section~\ref{simu}; this emission has a steeply falling spectrum due the the incoherence of the fields at these frequencies. The Cherenkov-like effect results in a highly forward-beamed emission. As seen previously, this electric field is polarized accordingly to the superposition of both mechanisms. Figure~\ref{selfas_allf}(left) presents the Fourier spectra of the simulated electric field for a vertical shower at $10^{17}$~eV and antennas located at different distances from the shower axis. For the antennas located at $100$ and $200$~m from the axis, we see the coherent part of the spectrum; this corresponds to the coherent geomagnetic and charge excess electric field between $30$~MHz and some hundreds of~MHz. The signal is incoherent at higher frequencies because these frequencies corrresponds to wavelength much smaller than the emittive zone.
Figure~\ref{selfas_allf}(right) shows the electric field amplitude between $300$~MHz and $1.2$~GHz as a function of the observer position at the ground level. The shower core is at the origin. At these frequencies, the Cherenkov-like ring clearly appears. This ring is due to the enhancement of the electric field when the particles velocity and the line of sight is close to the Cherenkov angle in air (see section~\ref{simu}).

The frequency domain was roughly separated in this paper between "below GHz" and "above GHz". This because another emission mechanism was studied and measured by Gorham {\em et al}.~\cite{PhysRevD.78.032007} in 2004 using a test beam at Argonne and SLAC: it is the molecular Bremsstrahlung radiation (MBR). This emission is due to the deceleration of the low-energy electrons ($\sim 10$~eV) in the plasma created by the high energy electrons and positrons of the shower. We remind that the electrons and positrons of the shower mainly loose energy by exciting or ionizing the medium. The excited N$_2$ then emits UV light detected by fluoresence telescopes but the ionized molecules provide low energy electrons forming a weakly ionized and stationary plasma. These electrons emit a Bremsstrahlung radiation that is expected to be non-polarized, isotropic which is potentially an excellent feature for a detection at large distance from the shower axis (the radiation is not focalized). The MBR implies a microwave continuum emission at the GHz level and is expected to have a direct relationship with the shower energy. Finally, this radiation is detectable with a duty cycle close to $100$\%.
The experiment of Gorham {\em et al}. measured a non-polarized flux density of $4\times 10^{-16}~\mathrm{W}~\mathrm{m}^{-2}~\mathrm{Hz}^{-1}$ for a setup corresponding to a shower at $3.36\times 10^{17}$~eV at $0.5$~m from the axis. Extrapolation to a typical observation distance of $10$~km leads to a threshold shower energy for a detection at $5\sigma$ of $1.6\times 10^{18}$~eV and $8.1\times 10^{18}$~eV assuming a quadratic or linear dependence of the power with respect to the shower energy respectively.
The MAYBE~\cite{2013EPJWC..5308008W} (electron beam) and AMY~\cite{amyICRC2013} (electron or positron beam) experiments also aimed at detecting and characterizing the MBR. The results disagree with those of Gorham. Recently, Conti et al.~\cite{2014arXiv1408.5886C}, using a low energy electron beam (in order to be below the Cherenkov threshold), obtained a linear scaling of the signal with the number of electrons and an asymmetric emission pattern, in opposition to the isotropic feature reported by Gorham {\em et al}.
In conclusion on the beam experiments, the situation is very unclear but the high MBR signal reported by Gorham {\em et al.} is not confirmed by any of the other experiments.

AMBER~\cite{PhysRevD.78.032007}, MIDAS~\cite{2013NIMPA.719...70A} and EASIER~\cite{ghzAugerICRC2013} are three prototypes installed at the Pierre Auger Observatory aiming at detecting the GHz emission from air showers. MIDAS excludes at quadratic (resp. linear) scaling of the power flux with the primary energy as reported in \cite{} at a confidence level of $5\sigma$ (resp. $4\sigma$). EASIER detected five (up to summer 2014) GHz events, in 3~years of data acquisition. All of those were detected by a single antenna and at a small distance from the shower axis (below $270$~m) and for high-energy showers. The simulation with SELFAS of the first detected EASIER shower was in good agreement with the data: it shows that the GHz signal can be explained with the usual geomagnetic and charge excess mechanisms in the GHz range, for this event. The others have not been simulated.

The CROME~\cite{CROME-ARENA2014} experiment, installed at the KASCADE experiment detected showers in the GHz range. The 35 detected showers in coincidence with the KASCADE setup after $18$~months of data taking presented an east-west asymmetry in their electric field strength.
High values of the electric field strength were measured following a ring pattern at the ground level. The detected GHz emission is strongly polarized with a pattern in agreement with the geomagnetic and charge excess expectations.
The conclusion of CROME is that the GHz signal associated to air showers is fully consistent with geomagnetic and charge excess mechanisms as sole emission processes at these frequencies and the data rule out the MBR mechanism.


\begin{table}[!ht]
\begin{center}
\begin{tabular}{|c|c|c|c|c|c|}\hline
\multicolumn{6}{|c|}{BEAM EXPERIMENTS}\\ \hline
Name & location & year & freq. (GHz)& scaling &  emission pattern \\ \hline
Gorham& SLAC&2004& $1$-$8$&quadratic  &isotropic \\ \hline
MAYBE &Argonne & 2012&$1$-$15$ & linear &isotropic, $\ll$ Gorham flux \\ \hline
AMY &Frascati &2012 &$1$-$20$ &\multicolumn{2}{|c|}{MBR much smaller than Cherenkov}\\ \hline
Conti et al.& Padova& 2014& $11$&linear & peaked forward\\ \hline
\multicolumn{6}{|c|}{PARABOLA IN CR EXPERIMENTS}\\ \hline
AMBER & Auger&  2011&\multicolumn{3}{|c|}{no CR detection}\\ \hline
MIDAS & Auger  & 2012&\multicolumn{3}{|c|}{no CR detection} \\ \hline
EASIER & Auger& 2011-2012& $3.4$-$4.2$&\multicolumn{2}{|c|}{\begin{minipage}{5cm}\vspace{0.2cm}5 CRs detected at high energy, very close to the shower axis\vspace{0.2cm}\end{minipage}} \\ \hline
CROME & KASCADE& 2011-2012& $3.4$-$4.2$&\multicolumn{2}{|c|}{\begin{minipage}{5cm}\vspace{0.2cm}35 CRs detected in agreement with geomagnetic and charge excess mechanisms, no MBR required\vspace{0.2cm}\end{minipage}}\\ \hline
\end{tabular}
\end{center}
\caption{Summary of the results obtained in the GHz domain, seeking for the characteristics of the molecular Bremsstrahlung radiation (MBR).}
\end{table}

In conclusion, the GHz signal strength and characteristics as reported in~\cite{PhysRevD.78.032007} are not confirmed by any other experiments. The GHz emission from showers is compatible with the GHz counterpart of the usual geomagnetic and charge excess mechanisms. No MBR emission associated with EAS is clearly observed up to now and clearly, detecting EAS using the GHz signal is not efficient.

\section{Radio signal below $\mathbf{20}$~MHz}
\begin{figure}[!ht]
\centerline{\includegraphics[scale=0.3,draft=false]{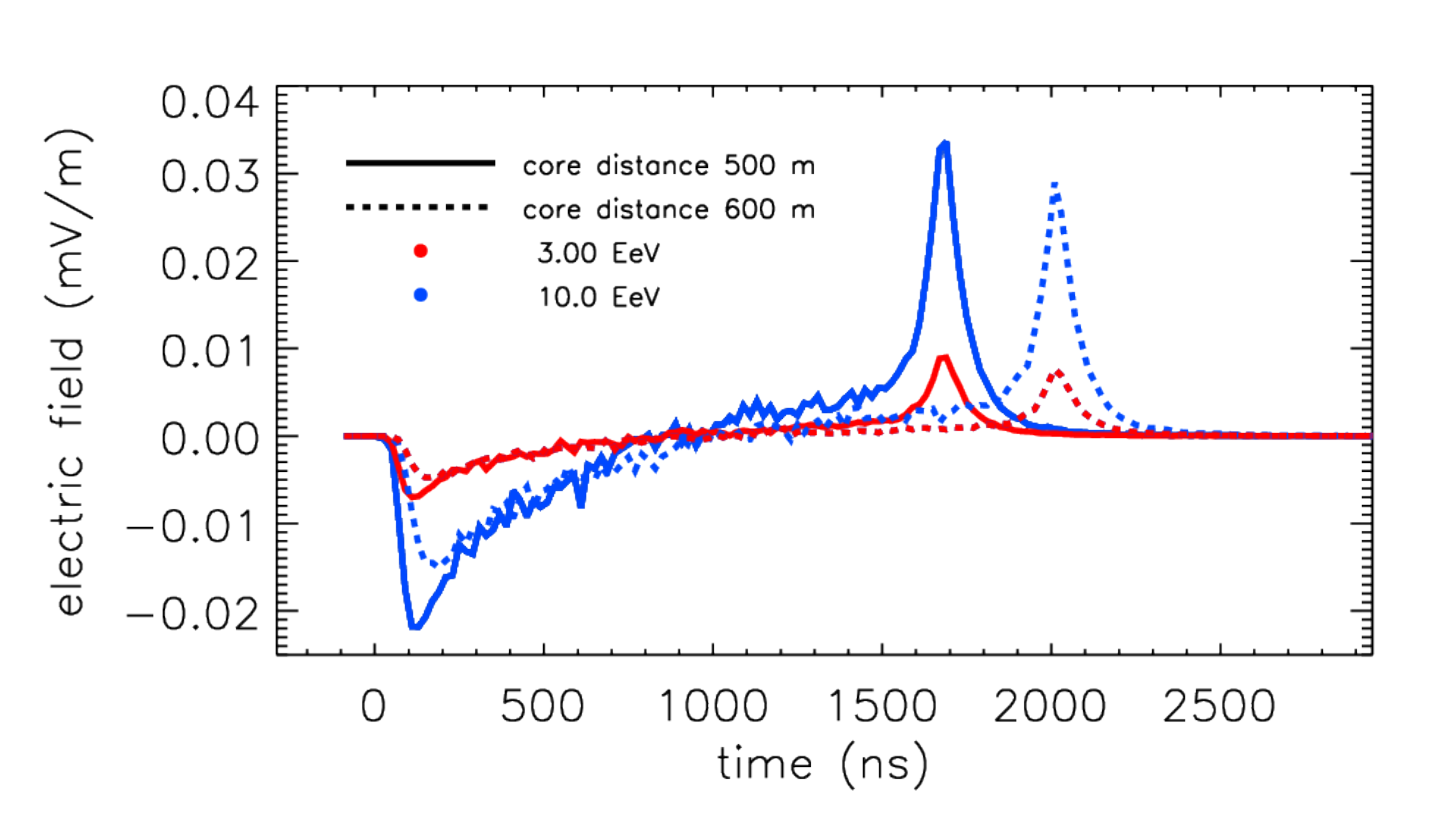}}
\caption{Vertical component of the electric field as a function of time for two observers located at $500$~m and $600$~m of the shower core for vertical simulated showers with primary energy of $3\times 10^{18}$~eV and $10^{19}$~eV. The electric field emitted during the shower development in the air is clearly visible around $t=100-200$~ns. The sudden death pulse, generated when the shower front hits the ground, appears at $t\sim 1700$~ns and $t\sim 2000$~ns for an observer at $500$~m and $600$~m, respectively, from the shower core.}\label{sdp}
\end{figure}

\begin{figure}[!ht]
\centerline{\includegraphics[scale=0.4,draft=false]{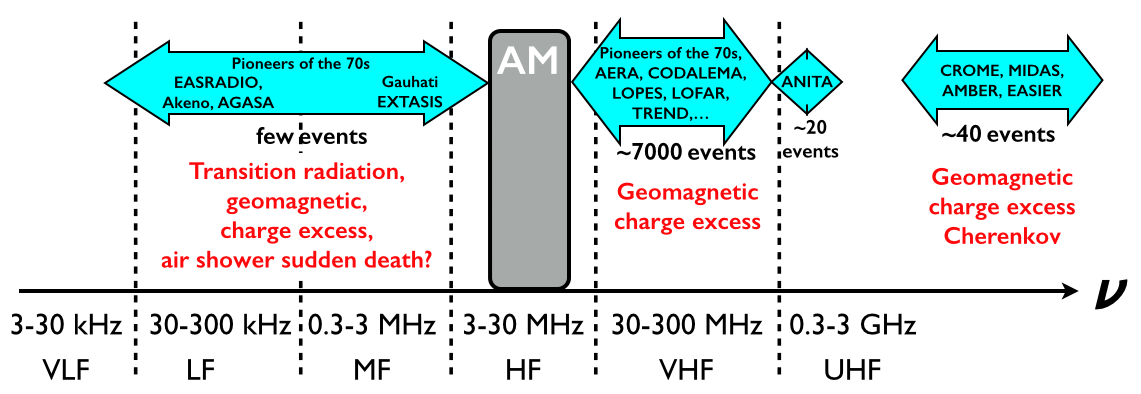}}
\caption[]{Summary of the current status of the experiments and of the understood mechanisms of electric field emission in showers, as a function of the frequency.}\label{freqtime}
\end{figure}

On the other side of the frequency spectrum, below $20$~MHz, we can of course think to the contributions of the geomagnetic and charge excess mechanisms. Nevertheless, these contributions have not been carefully measured at the MHz level because of the atmospheric noise which increases with decreasing frequency, this explains why people favoured higher frequencies. In the past, several experiments reported the observation of radio pulses in coincidence with EAS. A complete review of the experimental status can be found in Revenu~\& Marin~\cite{revenuicrc2013_sdp}. The frequencies probed at that time were between some hundreds of kHz up to $9$~MHz. The conclusions were that a very strong signal (1 or 2 orders of magnitude higher than in the usual band $30$-$80$~MHz) associated to air showers was undoubtly detected. People thought about several new mechanisms to explain this signal: the interaction between the ionization electrons in the air with the atmospheric electric field, the transition radiation front the electrons when the shower front hits the ground or the longitudinal and transverse emissions assuming full coherence but none of them could explain the reported huge values of the electric field. In~\cite{revenuicrc2013_sdp}, we propose a new emission mechanism: the coherent deceleration of the electrons and positrons of the shower front when it hits the ground level. The sudden deceleration of these particles when hitting the ground (which we call the sudden death of the shower) produces a coherent Bremsstrahlung emission for wavelengths larger than the typical size of the shower front, or equivalently for frequencies smaller than $20$~MHz. We can also understand this emission using a macroscopic point of view. When the shower front disappears below the ground level, the macroscopic charge density $\rho(\mathbf{r},t)$ and current $\mathbf{J}(\mathbf{r},t)$ vary very quickly and their space and time derivatives, providing electric fields according to Maxwell's equations, reach strong values. We therefore expect a significant electric field emission and we suspect that the previous experiments did measure this signal. We used the simulation code SELFAS to characterize the electric field emitted by the sudden death mechanism. The basic formula, derived from Eq.~\ref{seleq} and using the Coulomb gauge reads:
\be
\mathbf{E}_{tot}(\mathbf{r},t)=\frac{1}{4\pi\varepsilon_0 c}\frac{\partial}{\partial t}\sum _{i=1} ^{N_t}q_i(t_{\text{ret}})
\left[\frac{{\boldsymbol\beta}_{i}-(\mathbf{n}_i.\boldsymbol{\beta}_{i})\mathbf{n}_i}{R_i\,(1-\eta\boldsymbol{\beta}_i.\mathbf{n}_i)}\right]_{\text{ret}}\label{eq2}
\ee
where, as in Eq.~\ref{seleq}, $\eta$ is the air refractive index, $\mathbf{n}_i$ and $R_i$ are the line of sight and the distance between the observer and the particle $i$, $\boldsymbol{\beta}_i$ the velocity of this particle and $q_i$ its electric charge. The summation is done over the total number $N_t$ of particles that emitted an electric field detected by the observer at time~$t$.
All these quantities are evaluated at the retarded time $t_{\text{ret}}$, related to the observer's time $t$ through
$t=t_{\text{ret}}+\eta \,R_i(t_{\text{ret}})/c$.
Figure~\ref{sdp} presents the vertical component of the electric received by two observers located at $500$~m and $600$~m of the shower axis for vertical showers with a primary energy of $3\times 10^{18}$~eV and $10^{19}$~eV. The origin of time $t=0$ corresponds to the time when the shower front hits the ground.
In this figure, the electric field pulse created by the disappearance of the shower front in the ground has a very specific shape. Its maximum is located at a time corresponding to the time needed for the signal to reach the observer from the shower core. The reception time of this signal is therefore delayed by $t=d/c$ with respect to the time at the shower core, {\em ie} $1667$~ns at $500$~m and $2000$~ns at $600$~m. The shape of the pulse can be understood as the shape of the ground distribution of the electrons and positrons. The simulation shows that the sudden death pulse is polarized according to Eq.~\ref{eq2} in the direction of $\mathbf{a}-(\mathbf{n}\cdot\mathbf{a})\,\mathbf{n}$ because at first order, the $\boldsymbol{\beta_i}$ of the particles at the ground level is parallel to the shower axis $\mathbf{a}$. The vertical component should therefore be important. The sudden death pulse amplitude scales linearly with the primary energy and, a very important feature, is that this amplitude decreases as $1/d$ where $d$ is the distance to the shower core. It means that it should be possible to detect showers up to some kilometers from the core, contrarily to the signal in $30$-$80$~MHz.
Figure~\ref{freqtime} summarizes our current understanding of the emission mechanisms as a function of the frequency.

\section{Conclusion}
At the time of this conference, the radio signal emitted by extensive air showers is understood at a very high level of precision. This signal, between 30~MHz and 80~MHz, allows to reconstruct precisely the arrival direction of the primary cosmic ray. It also allows to estimate its nature following the results from the LOFAR experiment which claims a resolution of $20$~g~cm$^{-2}$. This value is comparable to the resolution of a fluorescence detector but with an uptime close to 100\% which is very interesting for improving our knowledge on ultra-high energy cosmic rays. The drawback of this signal is the relatively small range of the order of some hundreds of meters. The contribution from the geomagnetic and charge excess mechanisms can explain the data in a large range of frequencies, between 30~MHz up to some GHz. At low frequency, below 20~MHz, showers have already been observed in the past. A very large electric field amplitude was observed with no satisfactory underlying mechanism. We are currently installing the EXTASIS experiment in Nan\c{c}ay, dedicated to the low frequency signal of the showers when the front hits the ground. The range of this sudden death signal is much higher than the range in 30-80~MHz; this could be a very important feature for the detection of cosmic ray with a large efficiency.







\section*{References}
\bibliography{biblio,brevenu_spires,brevenu_proceedings_speaker,brevenu_proceedings,biblioComplete,biblioGroupe}

%
%
%


\end{document}